\documentstyle[sprocl]{article}

\input{psfig}
\def\Journal#1#2#3#4{{#1} {\bf #2}, #3 (#4)}
\def\NCA{\em Nuovo Cimento}

\def\PLB{{\em Phys. Lett.}  B}
\def\PRL{\em Phys. Rev. Lett.}
\def\PRD{{\em Phys. Rev.} D}

\def\be{\begin{equation}}
\def\ee{\end{equation}}
\def\bea{\begin{eqnarray}}
\def\eea{\end{eqnarray}}

\begin{document}

\title{PERTURBATIVE COLOR TRANSPARENCY IN ELECTRON BEAM
EXPERIMENTS VIA IMPACT-PARAMETER FACTORIZATION}

\author{JOHN P. RALSTON}

\address{
Department of Physics and Astronomy, University of Kansas, Lawrence,
KS 66045, USA\\ Email:ralston@ukans.edu }

\author{PANKAJ JAIN, BIJOY KUNDU}

\address{ Department of Physics, IIT Kanpur, Kanpur-208 016, India\\
E-mail: pkjain@iitk.ac.in, bijoyk@iopb.res.in}

\author{JIM SAMUELSSON}
\address{Department of Theoretical Physics,
University of Lund , Lund S-22362, Sweden\\ E-mail: jim@thep.lu.se}

\maketitle\abstracts{Color transparency is a prediction of
perturbative QCD. Yet detailed calculations have been lacking, and
aspects of the required factorization have been controversial.  We
report on the first complete calculations entirely within a
perturbative QCD framework. 
 We also
comment on the underlying factorization method and assumptions.}

\section{Introduction}

Color transparency is an
important tool to investigate exclusive processes using nuclear targets.
The earliest predictions of the phenomenon, going back to
1982,\cite{brodMuell} were based on having asymptotically large $Q^{2}$
select short distance.  Given the difficulties over short-distance
dominance, the transition to color transparency then multiplied the
controversy.  However the discovery of {\it nuclear filtering} shifted the
emphasis.\cite {JB,RP90,PBJ}  Components of a hadronic quark wave function
with large transverse spatial separations cannot propagate elastically in
nuclear matter.  This fact of QCD, noted earlier \cite {bertsch}
for diffractive processes, was re-discovered for color transparency. It was
realized that asymptotically large $Q^{2}$ is unnecessary to motivate
perturbative color transparency.

\medskip

Several experiments indicate that color
transparency and nuclear filtering have been observed at large nuclear
number $A$.  Perhaps
the most spectacular is the first color transparency experiment of
Carroll {\it et al},\cite {Car} which convincingly showed that
interference effects in proton-proton scattering were
filtered away in nuclear targets.  The FNAL E-665 experiment \cite {Fang}
also proved consistent with filtering effects,\cite {KNN93} especially in
the observation of longitudinal final state polarization in $\gamma^* A
\rightarrow \rho A$.  
 Electron beam experiments have been more difficult, partly
because of low
rates and reduced resolving power.  

\medskip

There are important theoretical reasons not to over-rely on the
$Q^{2}$ dependence of the transparency ratio.\cite {JainRalPRD}  The
ratio compares the object of study (an exclusive process in a nuclear
target) with something not well-understood (an analogous free-space
process).  The implicit assumption of a universal hard-scattering form
factor with fixed normalization and $Q^{2}$ dependence (to be followed
by some model of propagation) is theoretically unsupported.  Hybrid
models, in which free-space form factors are explicitly used, run into the
difficulty that exclusive processes are not self-consistently
described at laboratory energies by the asymptotic formalism.

\medskip
 Better methods exist
to characterize the data empirically. The $A$ dependence is particularly
powerful.  O'Neill
 {\it et al} \cite {Neill} showed that effective attenuation cross sections
extracted from $ A (e, e'p)$ SLAC data were smaller than Glauber theory
calculations by a statistically significant amount. However, choice of
normalization and uncertainties in the nuclear spectral distributions
complicated the interpretation, and the precision of the data \cite{Mak} was
insufficient to establish a large effect. New $(e, e'p)$ beam experiments
are underway at Jefferson Laboratory.

\medskip

The theory to describe color transparency in pQCD actually forced a
revision of basic factorization methods.\cite{RP90}  It was found that
the asymptotic factorization 
of Lepage and Brodsky \cite {GPL} ($LB$) 
 was inadequate.  To go beyond this, it was necessary to
introduce an integration over the transverse separation of quarks into
the description.  An antecedent was an important 
paper by Botts and Sterman.\cite{Botts}  The promising new
factorization methods, which might be called ``impact parameter
factorization", have been highly refined in applications by Li and
Sterman \cite{LS,L} to free-space form factors.  Kundu
{\it et al} \cite {KLSJ} developed the method further, and responded to
criticisms.\cite {JKB}

\medskip

The purpose of the transverse integrations is
to incorporate regions of finite quark spatial separation. In
free-space the Sudakov form factor tends
to suppress these regions somewhat; in nuclei
a stronger suppression comes from the filtering
effects of nuclear matter.\cite {us}  The new method also enables
calculation of
helicity-flip form factors,\cite {BNL93} which cannot be
described in the old formalism, and which
would be very interesting to measure in nuclear matter, if somehow it
could be done.

\medskip

Happily one does not give up the impulse approximation, which has been
the cornerstone of all successful pQCD calculations.  Often
misconstrued to be the same as the ``frozen approximation'' of
high-energy diffractive processes, the impulse approximation allows
one to start a clock allowing time evolution of the outgoing system,
while separating the fast time scale of the scattering from the slow
time scale of hadron formation.  Controversies in the literature still
rage over whether ``expansion'' in one form or the other has been
incorporated properly.\cite {PBJ}  Our calculations in perturbation
theory integrate over light-cone ``minus'' components of
wavefunctions.  This is the step which separates the hadron time scale
from the rest.  The rest of the calculation is Feynman diagrams, which
have the time evolution of quarks built in.  Feynman diagrams faithfully
reproduce time-evolution of the quark and gluon degrees of freedom
order-by-order, creating the same kind of amplitude (a Green function) that
the experiment can measure.  The physical picture of the impact-parameter
factorization, then, is somewhat different from the picture of the asymptotic
limit.  In the asymptotic limit, quarks participate at zero distance,
and move so fast that perfect transparency always occurs.  In
impact-parameter factorization, quarks with all sideways and
longitudinal separations
are superposed coherently over the whole nucleus.  The hard scattering
and the propagation remain coupled during propagation.  

\section{The Method}

It is worthwhile to review the different frameworks of exclusive
processes in free space before introducing nuclear targets.
Lepage and Brodsky \cite{GPL} ($LB$) calculate a meson electromagnetic form
factor with a factorization
written as
\begin{equation}
F_\pi(Q^2) = \int dx_1 dx_2 \phi(x_2,Q) H(x_1,x_2,Q) \phi(x_1,Q).
\end{equation}
Here $\phi(x,Q) $ are the distribution amplitudes, which can be
expressed in terms of the pion wave function $\psi(x,\vec k_T)$ as
$ \phi(x,Q) = \int^Q d^2k_T\psi(x,\vec k_T)$.
We use $x$ for the longitudinal momentum fraction and $\vec k_T$ for
the transverse momentum carried by the quark.  The factorization is
justified provided the external photon momentum $Q^2$ is
{\it asymptotically large}.  Then the $k_T$ integrals of perturbation
theory decouple, and can be applied to make the distribution amplitudes. The
$k_T$ dependence of the hard scattering $H$ can be expanded in a power
series, retaining the trivial, constant term.  One directly obtains the
power-law scaling of the quark-counting method,\cite{GRF} with logarithmic
corrections.

\medskip

Note that the asymptotic limit is taken right away, in fact
prematurely. All distances separating quarks then become asymptotically
short.  For the purposes of color transparency, taking $Q^{2}$ arbitrarily
large (but fixed), one might think all targets become perfectly
transparent. But then taking $A \rightarrow \infty$ at fixed, arbitrarily
large $Q^{2}$,  we know that all targets must become opaque.  Thus there is a
limit interchange problem in the $LB$ factorization, because the limit of
large $Q^{2}$ and large $A$ do not commute. The scheme is fundamentally
limited to asymptotic $Q^{2}$, and there is no way to fix it to describe
the phenomena of color transparency at laboratory energies.

\medskip

A proper description of the phenomena follows from a impact-parameter
factorization scheme incorporating the transverse degrees of 
freedom.\cite{RP90,Botts}  By including a broader integration region,
impact-parameter factorization is more general than the $LB$ method.
Li and Sterman \cite{LS} simplify the calculation of form-factors by
dropping the weak $k_T$ dependence of quark propagators in a hard
scattering kernel $H$.  
Working in configuration (impact-parameter $b$) space 
the expression for a form-factor
becomes: \begin{equation} F_\pi(Q^2) = \int dx_1 dx_2 {d^2\vec b\over
(2\pi)^2} {\cal P}(x_2,b,P_2,\mu) \tilde H(x_1,x_2,Q^2,\vec b,\mu)
{\cal P}(x_1,b,P_1,\mu), \end{equation} where ${\cal P}(x,b,P,\mu)$
and $\tilde H(x_1,x_2,Q^2,\vec b,\mu)$ are the Fourier transforms of
the wave function, including Sudakov factors, and hard scattering
respectively; $\vec b$ is conjugate to $\vec k_{T1} - \vec k_{T2}$,
$\mu$ is the renormalization scale and $P_1$, $P_2$ are the initial
and final momenta of the meson.

\medskip

We now discuss filtering.  When a fast hadron traverse the nuclear
medium, each quark interacts primarily through exchange of transverse
momentum.  While longitudinal momentum can also be exchanged, this
degree of freedom does not affect the overall structure, and is in fact
lost in the present uncertainties of Feynman $x$ wave-functions.
Nuclear targets provide a much better transverse filter than free
space, and so it was thought at first that the $LB$ method might
apply.  The profound differences between impact-parameter
factorization and the $LB$ factorization was not understood at first.
It was thought that the Ralston and Pire 1990 \cite{RP90} method was a way to
generate distribution amplitudes within the $LB$ framework.
Eventually we realized that the large $Q^{2}$ dependence of the
impact-parameter method is vastly more flexible than just logarithmic
behavior, while $LB$ factorization cannot be otherwise.

\medskip Transverse momentum integrations conserving overall momentum,
and at small momentum transfer, turn into products in $b$ space.  Thus
the nuclear medium modifies the quark wave function such 
that \cite{RP90}
${\cal P}_A(x,b,P,\mu) = f_A(b; B){\cal P}(x,b,P,\mu) $,
where ${\cal P}_A$ is the wave function probed inside the medium and
$f_A$ is the nuclear filtering amplitude. 
 An eikonal form \cite{RP90}
appropriate for $f_A$ is: $f_A(b; B) = exp(-\int_{z}^{\infty} dz'
\sigma(b) \rho(B, z')/2) $.  Here $\rho(B, z')$ is the nuclear number
density at longitudinal distance $z'$ and impact parameter $B$ relative
to the nuclear center.  
We parametrize $\sigma(b)$ as $k b^2$ for our calculations.
Finally, we must include the probability to find a target at position
$B, z$ inside the nucleus.  Putting together the factors, the process
of knocking out a hadron from inside a nuclear target has an amplitude
$M$ given by

\begin{eqnarray}
M &=& \int_{0}^{\infty} d^2 B \int_{-\infty}^{+\infty} dz \rho(B,z)
\times F_{\pi}(x_1,x_2,b,Q^2) \times f_A(b,B)
\end{eqnarray}

For the proton the important transverse scale is the maximum of the
three quark separation distances, $b_{max} = max(b_1,b_2,b_3)$.
The calculation of the process in the nuclear target needs a 9
dimensional integration, which is performed by Monte Carlo.
The calculation of proton targets, then, follows the same basic rules as
knocking out pions, but with more degrees of freedom.  In our calculations
we found that uncertainties on the nuclear correlations at the $10\% $
level were a major concern, in some cases exceeding the theoretical
uncertainties from the rest of the calculation.\cite {us}
We also find that
 the physics cannot
reasonably be captured by a free-space hard scattering, followed by some
model of propagation with or without ``expansion''.  This is because the
integrations over the transverse quark variable extend over the whole
volume of the nucleus. Color transparency
is something probing the internal structure of hadrons.

\section {Results and Discussions}

 We calculated color transparency and nuclear
filtering for populations of both pions and nucleons (protons) in the
nucleus.  We explored the $Q^{2}$ dependence of the proton
transparency ratio using popular distribution amplitude models, and
later (after the meeting) with the asymptotic and other models.  
 Realizing belatedly
that wave functions
more central in Feynman $x$ would possibly be more purely
short-distance, we explored such cases and include them for comparison.
To make our
trial
transparency ratio, we divided the rate in the nuclear
target by a cross section including the free-space
form factor.

\begin{figure} [t,b] \hbox{\hspace{6em}
\hbox{\psfig{figure=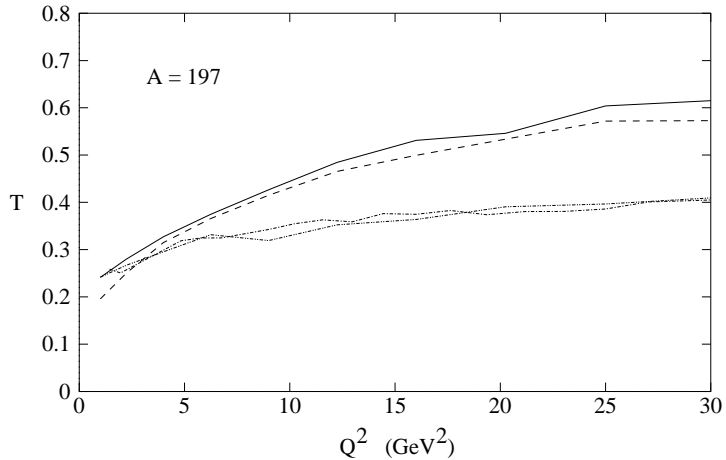,height=6cm}}} \caption{
Color transparency ratio for different proton trial wave functions.
Solid curve: asymptotic wave function. Dashed curve: a more centrally
peaked wave function of the type
$x_1^2*x_2^2*x_3^2$. Dash-dot curve: $KS$ wave
function. Dash-dot-dot curve: $CZ$ wave
function.} \label{fig1}
\end{figure}

Some results are shown in Fig.  \ref{fig1}.  It is noteworthy that the
{\it slope} of the ratio versus $Q^{2}$ depends on the
wave-functions. The ratio for a model having a more-central
wave function rises faster than that of model with large endpoint
contributions.  This
is consistent with the known tendency for endpoint regions of the $CZ$
and $KS$ models \cite {CZ} to enhance soft regions of integration, or
``big-fat''
protons.  The asymptotic model, for example, is seen to look like a
``smaller'' proton in the calculation with the quarks sitting in the
central regions.

\medskip

We are not concerned here with controversies over different trial wave
functions, and use the asymptotic one merely to
illustrate a point.  It is quite striking, and possibly important, that the
$Q^{2}$
dependence of the transparency ratio measures something as fundamental
as the basic $x$ depedence of the wave functions. The calculations
indicate that the slope can distinguish
centrally peaked from endpoint-peaked models.

\medskip
For the endpoint-dominated models, the quark transverse separation cutoff
showed a significant reduction in
sensitivity inside the nucleus compared to free space. 
 However, it must
be remembered that the
effects of filtering in electron beam experiments are comparatively
modest compared to hadron-hadron reactions, because the hadron tends
to be knocked out the ``back-side'' of the nucleus. This accounts for
an intrinsic reduced sensitivity of electron beam reactions, which
hopefully is compensated by their high intrinsic precision.

\medskip

\begin{figure} [t,b] \hbox{\hspace{6em}
\hbox{\psfig{figure=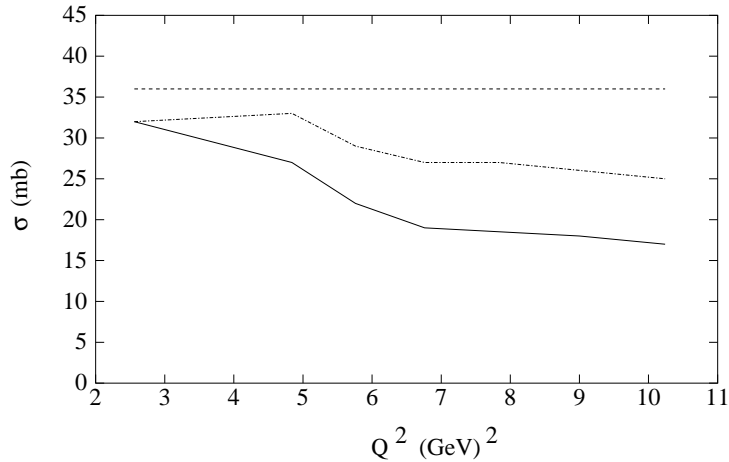,height=6cm}}}
\bigskip
\caption{
Extracted effective attenuation cross sections $\sigma_{eff} (Q^2)$ as a
function of $Q^2$ exhibit color transparency. The decrease of $\sigma_{eff}
(Q^2)$ with $Q^2$ is sufficiently large that conventional nuclear physics
might be ruled out with large $Q^2$ or sufficient precision. Solid curve:
asymptotic
wavefunction. Dash-dot curve: KS wave function. The
dashed line is the Glauber value of 36 mb. } \label{fig2} \end{figure}

Finally, following,\cite{JainRalPRD} we extracted the effective
attenuation cross section $\sigma_{eff} (Q^2)$, which serve as a litmus
test of whether ``color transparency" has actually been achieved. 
The
results (Fig. \ref{fig2}) show a significant decrease of $\sigma_{eff}
(Q^2)$ with increasing $Q^2$ to values well below the Glauber model
attenuation cross section. This indicates color transparency.
Consistently, the transparency of the asymptotic model is
more dramatic than the cases of the $CZ$ and $KS$ models.

\medskip

Uncertainties in nuclear spectral functions remain, and are
becoming problematic for calculation of absolute normalizations.  This
strongly affects the $Q^{2}$ dependence, but is somewhat ameliorated
by studying the nuclear matter limit of large $A$. Several conclusions are
nevertheless well supported by the investigations,
reported in more detail elsewhere.\cite {us}   These first
quantitative perturbative calculations support the claim that
exclusive processes are theoretically under better control than the
free-space analogues.  Next, observables such as the slope of the
transparency ratio, sometimes calculated in terms of an ad-hoc
``expansion time scale'' in hadronic models, probe the interplay of the
transverse and longitudinal
development of the amplitudes in perturbation theory. The slope
depends directly on the
$x$-dependence of wave functions.
Next, the calculations show significant reduction in attenuation, even
in circumstances where {\it no rise is seen with $Q^2$ in
transparency ratios}.  The explanation, we believe, is that
purification to
short-distance tends to deplete amplitude normalizations. Division by
a free-space cross section, where uncontrolled amplitudes
dominate, overestimates the magnitude of hard scattering inside a
nucleus, and can give a misleading ratio. This competition is
unlikely to be unraveled with $Q^{2}$ dependence alone, and calls for
another experimetal handle.  The $A$ dependence separates the
competition between process normalization and attenuation, as revealed
by $\sigma_{eff}$.

\section *{Acknowledgments}
This work was supported by
BRNS grant No. DAE/PHY/96152,
 the Crafoord Foundation, the Helge Ax:son Johnson Foundation, DOE Grant
 number
DE-FGO2-98ER41079,
the KU General Research Fund and NSF-K*STAR
Program under the Kansas Institute for Theoretical and Computational
Science.

\section *{References}
 \end{document}